# Ultrahard BC$_5$ - an efficient nanoscale heat conductor through dominant contribution of optical phonons

*Rajmohan Muthaiah[1], Jivtesh Garg[1], Shamsul Arafin[2]

[1]School of Aerospace and Mechanical Engineering, University of Oklahoma, Norman, OK-73019, USA

[2]Department of Electrical and Computer Engineering, Ohio State University, Columbus, OH-43210, USA

**Abstract:** In this work, we study thermal conductivity ($k$) of BC$_5$, an ultra-hard diamondlike semiconductor material, using first-principles computations and analyze the effect of both isotopic disorder as well as length scale dependence. $k$ of isotopically pure BC$_5$ is computed to be 169 Wm$^{-1}$K$^{-1}$ (along a-axis) at 300K; this high $k$ is found to be due to the high frequencies and phonon group velocities of both acoustic and optical phonons owing to the light atomic mass of Carbon (C) and Boron (B) atoms and strong C-C and B-C bonds. We also observe a dominance of optical phonons (~ 54%) over acoustic phonons in heat conduction at higher temperatures (~500 K). This unusually high contribution of optical phonons is found to be due to a unique effect in BC$_5$ related to a weaker temperature dependence of optical phonon scattering rates relative to acoustic phonons. The effect is explained in terms of high frequencies of optical phonons causing decay into other high frequency phonons, where low phonon populations cause the decay term to become insensitive to temperature. The effect further leads to high nanoscale thermal conductivity of 77 Wm$^{-1}$K$^{-1}$ at 100 nm length scale due to optical phonon meanfreepaths being in nanometer regime. These results provide avenues for application of BC$_5$ in nanoscale thermal management.

**Keywords:** Boron, Carbon, thermal conductivity, optical phonons, BC$_5$, ultra-hard materials, thermal management, phonons, nanoelectronics

**Introduction:** Materials with high thermal conductivity are critical for efficient heat dissipation in power electronics and electronics cooling. Boron and carbon based compound-semiconductors are promising materials for thermal management due to their high thermal conductivity emerging from light mass of atoms involved and strong bonds of C-C and B-C[1-6]. An ultra-high thermal conductivity of 2305 Wm$^{-1}$K$^{-1}$ for the super-hard bulk hexagonal BC$_2$N was reported by Safoura

*Corresponding author. Tel: 409 665-1351. E-mail: rajumenr@ou.edu (Rajmohan Muthaiah)

*et al.*[7]. Similarly, high anisotropic thermal conductivities of 1275.79 Wm$^{-1}$K$^{-1}$ and 893.90 Wm$^{-1}$K$^{-1}$ were reported for monolayer BC$_2$N along zig-zag and arm-chair directions[8]. BC$_5$ is a diamond-like ultra-hard semiconductor with exceptional hardness of 83 GPa and experimental bulk modulus of 335 GPa[9, 10]. In this work we use first principles calculations to analyze the thermal conductivity of BC$_5$. At 300 K, we report a high thermal conductivity ($k$) of 169 Wm$^{-1}$K$^{-1}$ for bulk BC$_5$ (inifinte dimensions) along *a*-axis. A high nanoscale thermal conductivity of ~ 51 Wm$^{-1}$K$^{-1}$ is reported for length scale of 50 nm (at 300 K), indicating BC$_5$ will be a promising material for thermal management in nanoelectronics. To understand the origin of this high nanoscale thermal conductivity we systematically analyzed elastic constants, phonon group velocities and phonon scattering rates of different phonon modes. The high nanoscale thermal conductivity is found to be due to a dominant contribution of optical phononos to overall thermal conductivity in BC$_5$; at 500 K and 1000 K, optical phonons (with meanfreepaths in the nanometer regime) contribute almost ~54% and 57.3%, respectively to the overall thermal conductivity along the a-axis. First-principles computations are used to shed light on the dominant role of optical phonons in BC$_5$ thermal conductivity.

**Computational methods:** First principles computations were performed using local density approximations[11] with norm-conserving pseudopotentials using QUANTUM ESPRESSO[12] package. The geometry of the tetragonal (space group P3m1) BC$_5$ with 6 atoms unit cell is optimized until the forces on all atoms are less than 10$^{-5}$ eV Å$^{-1}$ and the energy difference is converged to 10$^{-12}$ Ry. A plane-wave cutoff energy of 100 Ry was used. Electronic calculations were performed using 12 x 12 x 6 Monkhorst-Pack[13] *k*-point mesh. Optimized BC$_5$ structure is shown in Fig. 1; obtained lattice constants of a=2.516 Å and c/a=2.506 are in good agreement with the previous first principles calculations[14]. Elastic constants were computed using QUANTUM ESPRESSO thermo_pw package and Voigt-Reuss-Hill approximation[15] is used to calculate the bulk modulus, shear modulus(G) and Young's

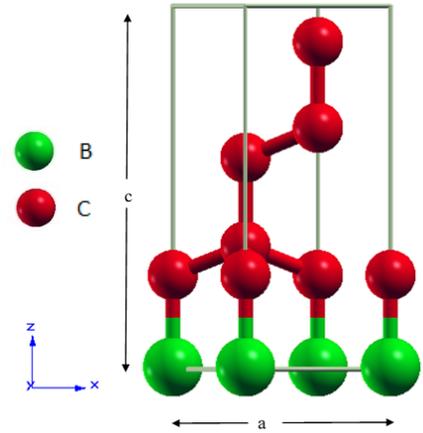

Figure 1: Atomic arrangements of BC$_5$ with space group P3m1. Red and green sphere represents the carbon and boron respectively.

Modulus(E). Lattice thermal conductivity is calculated by deriving the most important ingredients, namely, the harmonic and anharmonic interatomic force interactions from density-functional theory and using them with an exact solution of the phonon Boltzmann transport equation (PBTE)[16-18]. Thermal conductivity ($k$) in the single mode relaxation time (SMRT) approximation[19] (usually the first iteration in solution of PBTE) is given by,

$$k_\alpha = \frac{\hbar^2}{N\Omega k_b T^2} \sum_\lambda v_{\alpha\lambda}^2 \omega_\lambda^2 \bar{n}_\lambda (\bar{n}_\lambda + 1) \tau_\lambda \qquad (1)$$

where, $\alpha$, $\hbar$, N, $\Omega$, $k_b$, T, are the cartesian direction, Planck constant, size of the q mesh, unit cell volume, Boltzmann constant, and absolute temperature respectively. $\lambda$ represents the vibrational mode ($qj$) ($q$ is the wave vector and $j$ represents phonon polarization). $\omega_\lambda$, $\bar{n}_\lambda$, and $v_{\alpha\lambda}$ ($= \partial\omega_\lambda/\partial q$) are the phonon frequency, equilibrium Bose-Einstein population and group velocity along cartesian direction $\alpha$, respectively of a phonon mode $\lambda$. $\omega_\lambda$, $\bar{n}_\lambda$, and $c_{\alpha\lambda}$ are derived from the knowledge of phonon dispersion computed using 2$^{nd}$ order IFCs. $\tau_\lambda$ is the phonon and is computed using the following equation,

$$\frac{1}{\tau_\lambda} = \pi \sum_{\lambda'\lambda''} |V_3(-\lambda, \lambda', \lambda'')|^2 \times [2(n_{\lambda'} - n_{\lambda''})\delta(\omega(\lambda) + \omega(\lambda') - \omega(\lambda'')) + (1 + n_{\lambda'} + n_{\lambda''})\delta(\omega(\lambda) - \omega(\lambda') - \omega(\lambda''))] \qquad (2)$$

where, $\frac{1}{\tau_\lambda}$ is the anharmonic scattering rate due to intrinsic three phonon interactions and $V_3(-\lambda, \lambda', \lambda'')$ are the three-phonon coupling matrix elements computed using both harmonic and anharmonic interatomic force constants. Dynamical matrix and harmonic force constants were calculated using 8 x 8 x 4 q-grid. 4 x 4 x 2 q-points were used to compute the anharmonic force constants using QUANTUM ESPRESSO D3Q[16, 18, 20] package. Acoustic sum rules were imposed on both harmonic and anharmonic interatomic force constants. Phonon linewidth and lattice thermal conductivity were calculated iteratively using QUANTUM ESPRESSO thermal2 code with 30 x 30 x 15 **q**-mesh and 0.05 cm$^{-1}$ smearing until the $\Delta k$ values are converged to 1.0e$^{-5}$. Casimir scattering[21] is imposed for length dependence thermal conductivity calculations.

**Results and discussions:**

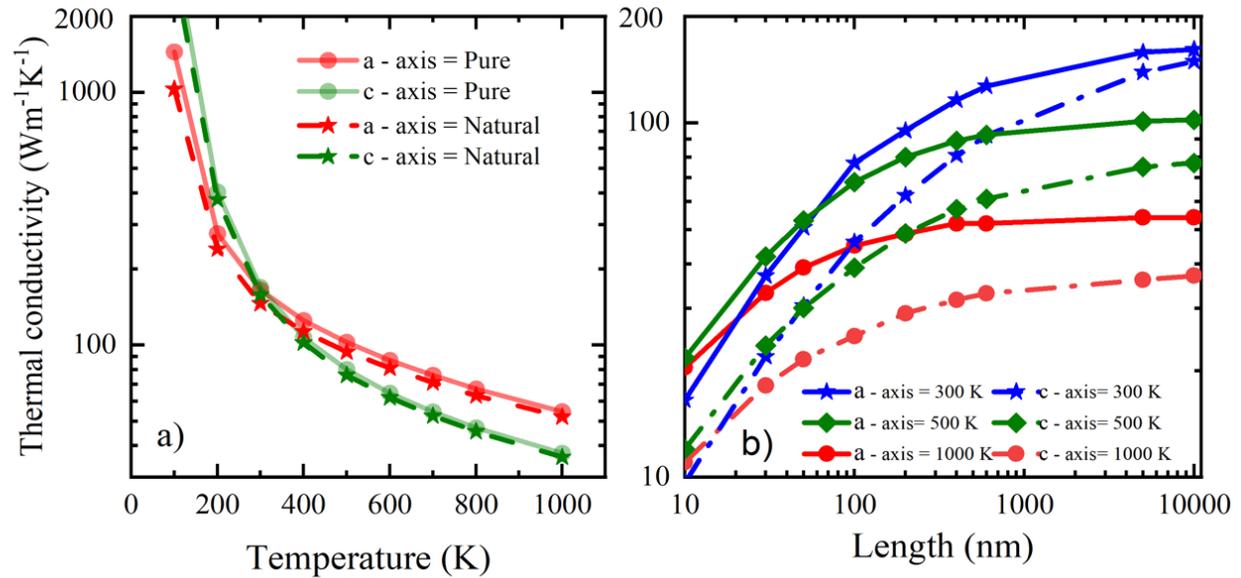

Figure 2: a) Thermal conductivity of $BC_5$ along a-axis and c-axis at different temperatures b) Length dependence of thermal conducticity (300 K) of $BC_5$ between 10 nm and 10000 nm.

Lattice thermal conductivity of isotopically pure and isotopically disordered (naturally occurring) $BC_5$ along a-axis and c-axis is shown in Fig 2a as a function of temperature. At room temperature (300 K), computed thermal conductivities of bulk $BC_5$ along a-axis and c-axis are 165 W/mK and 169 $Wm^{-1}K^{-1}$, respectively, for pure $BC_5$. Our reported $k$ values are higher than the $k$ values of silicon (153 $Wm^{-1}K^{-1}$)[22] at room temperature suggesting that $BC_5$ will be a promising material for thermal management applications.

      This high thermal conductivity of $BC_5$ is a direct consequence of high phonon frequencies and phonon group velocities in BC5 as seen in Fig. 3 resulting from the strong C-C and B-C bonds, and the light mass of B and C atoms. While in silicon, the maximum LA phonon frequency reaches 400 $cm^{-1}$ [23], Fig. 3 shows that in $BC_5$, LA phonons have higher frequencies, reaching values greater than 600 $cm^{-1}$. The strong bonding in $BC_5$ is seen by noticing that the bulk modulus and Young Modulus of $BC_5$ are higher than Silicon and almost equal to the values for Diamond (computations for Diamond and Silicon are discussed in supplementary section).This is due to the strong covalent bond network through *sp³* hybridization[24].

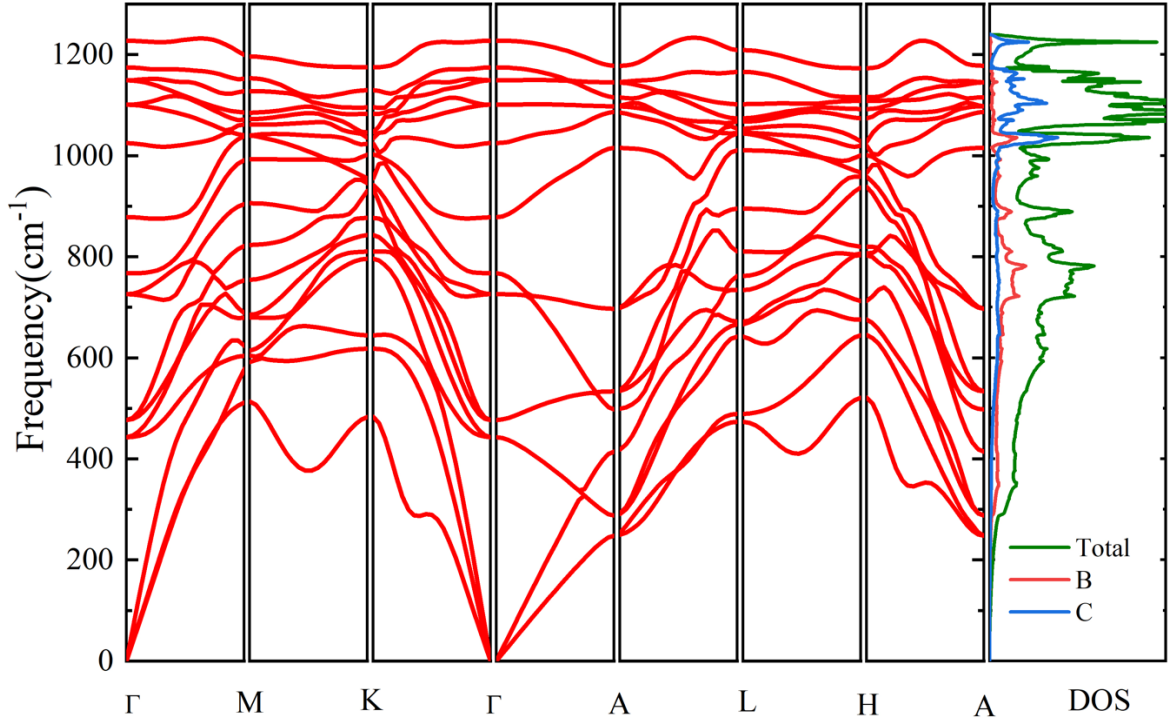

Figure 3: Phonon dispersion along the high symmetry points of tetragonal $BC_5$ and and phonon density of states

We also estimated $k$ of naturally occurring $BC_5$ to be 146 W/mK and 158 Wm$^{-1}$K$^{-1}$, along a-axis and c-axis, respectively at 300 K. $k$ of naturally occurring $BC_5$ is only lower by 11.5% and 6.5% relative to pure $BC_5$ along a-axis and c-axis respectively. Thermal conductivity for naturally occurring $BC_5$ was computed by introducing additional phonon scattering arising out of mass-disorder due to random distribution of isotopes of Boron and Carbon throughout the crystal. The small mass variation in isotopes of both B (atomic mass of 10.013 a.u with 19.9% concentration and atomic mass of 11.009 a.u with 80.1% concentration) and C (atomic mass of 12 a.u with 98.93% concentration and 13.003 a.u with 1.07% concentration) atoms[25], induces only a small additional phonon scattering, causing only a minor decrease in thermal conductivity of naturally occurring $BC_5$ relative to the pure case.

**Table 1: Elastic constants of BC5, Silicon and Diamond in GPa**

| Material | C11 | C33 | C44 | C66 | C12 | C13 | Bulk Modulus | Young modulus | Shear Modulus |
|---|---|---|---|---|---|---|---|---|---|
| $BC_5$ | 911 | 1061.8 | 394.5 | 361.1 | 189 | 97 | 405 | 894 | 396 |

| | | | | | | | | |
|---|---|---|---|---|---|---|---|---|
| Silicon | 159.5 | 159.5 | 78.1 | 78.1 | 61.3 | 61.3 | 94.1 | 158.1 | 64.8 |
| Diamond | 1099.5 | 1099.5 | 601.4 | 601.4 | 127.1 | 127.1 | 451.3 | 1176.8 | 552.3 |

Length dependent $k$ of pure $BC_5$ was also investigated by introducing Casimir scattering (boundary scattering). High thermal conductivity of ~ 51 $Wm^{-1}K^{-1}$ at nanometer length scale of L = 50 nm (at 300 K) is observed in Fig. 2b. Nanoscale thermal conductivity of $BC_5$ is more than a factor of 2 higher than silicon as seen in Fig. 4 where we compare the phonon meanfreepath dependence of thermal conductivity accumulation in $BC_5$ and silicon. In silicon, phonons with meanfreepath below 100 nm contribute only ~40 W/mK at 300 K; in $BC_5$, however, phonons in the same meanfreepath range, contribute a much higher value of ~95 W/mK along a-axis in BC$_5$. This higher nanoscale thermal conductivity of BC$_5$ provides promising new avenues for achieving efficient nanoscale thermal management.

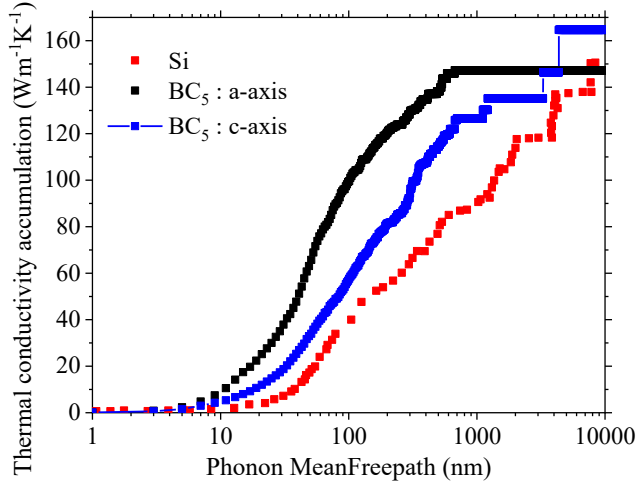

Figure 4: Phonon meanfreepath dependence thermal conductivity of BC5 and Silicon at 300 K

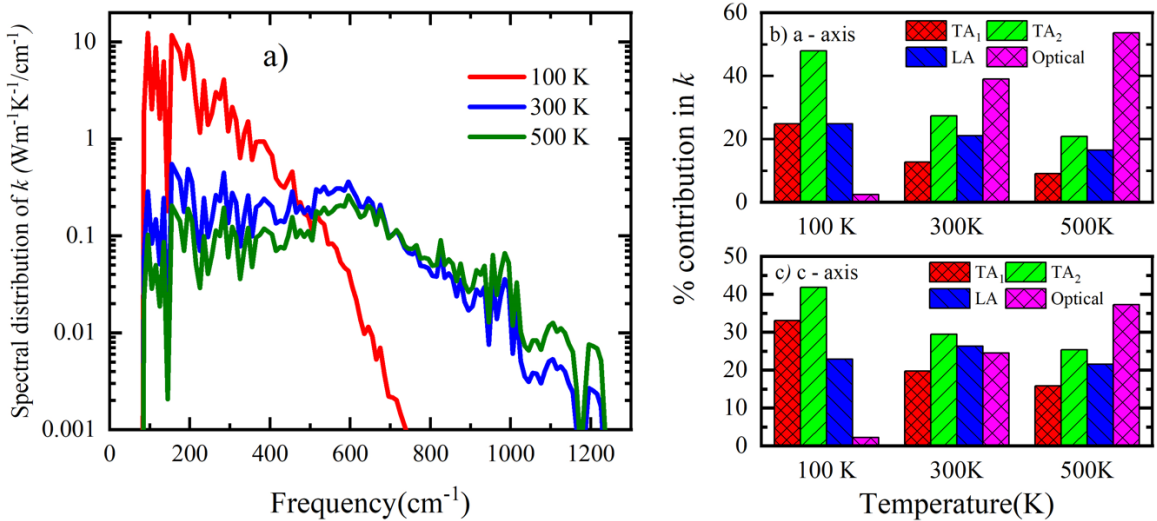

Figure 5a) Spectral distribution of $k$ with frequency at 100 K, 300 K and 500 K. b and c) Percentage contribution by transverse acoustic ($TA_1$ and $TA_2$), longitudinal acoustic (LA) and optical phonon modes.

This much higher nanoscale thermal conductivity in BC$_5$ is found to be due to the dominant role played by optical phonons in conducting heat in BC$_5$. At 500 K and 1000 K, optical phonons contribute 54% and 57.3%, respectively, to overall $k$ along a-axis. This is in contrast to typical semiconductors like silicon, where optical phonon contribution to $k$ is in the range of ~5%[26] at 300 K. Large contribution of optical phonons can also be seen in Figures 5a and b which show the spectral distribution of $k$ as well as percentage contribution of transverse acoustic (TA), longitudinal acoustic (LA) and optical phonon modes to overall $k$. At T > 300 K, optical phonons have a considerable contribution to overall thermal conductivity.

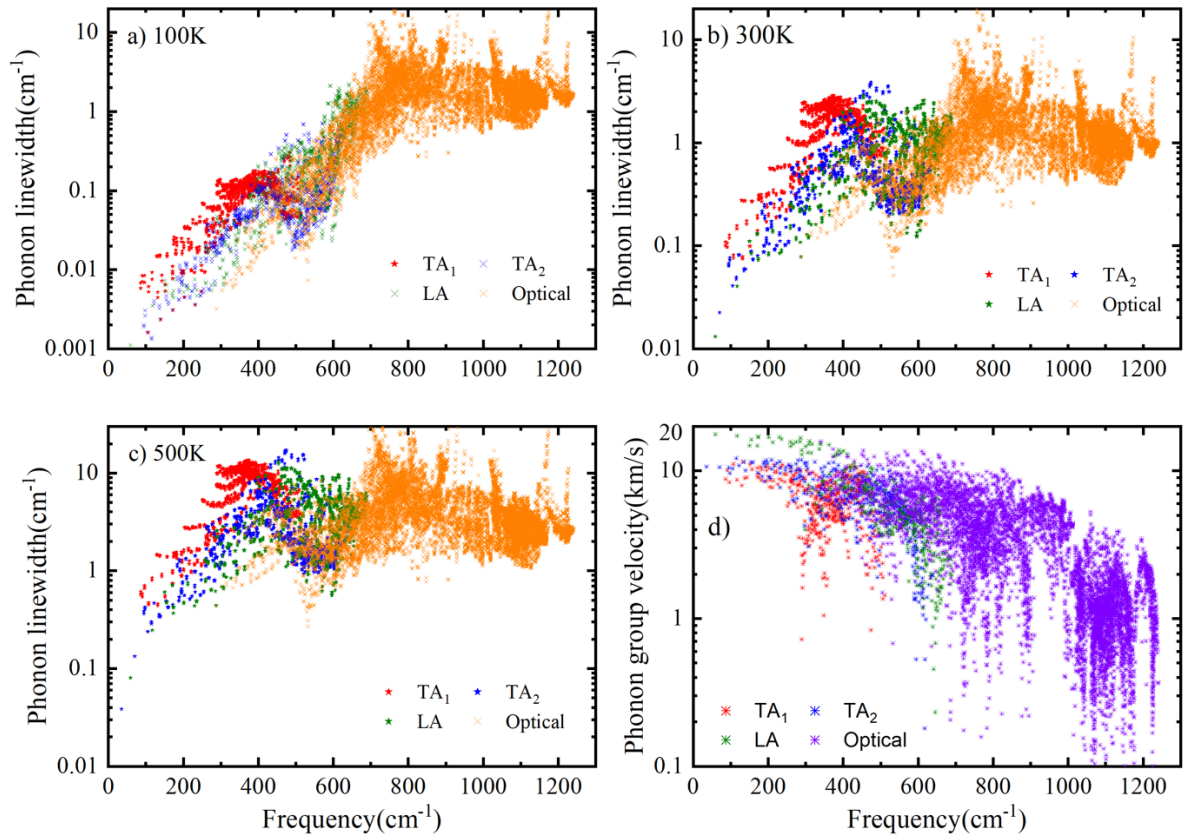

Figure 6 Phonon linewidths of transverse acoustic, longitudinal acoustic and optical phonon modes at a) 100 K b) 300 K c) 500 K. d) Phonon group velocity of all phonon modes

First-principles computations reveal that this dominant contribution of optical phonons to overall thermal conductivity in $BC_5$ is due to a combination of high optical phonon velocities (Fig. 6d) and comparable optical phonon scattering rates to acoustic phonons at temperatures greater than 300 K (Fig. 6a-c). This second effect is particularly interesting, since higher optical phonon frequencies typically result in optical phonon scattering rates to be significantly larger than acoustic phonons. While at 100 K, optical phonon scattering rates in $BC_5$ are indeed higher than acoustic phonons (Fig. 6a), as temperature increases optical-phonon scattering rates increase at a much slower rate compared to acoustic phonons, causing optical and acoustic phonon scattering rates to become comparable (Fig. 6b and c).

This weak dependence of optical phonon scattering rates on temperature and comparable scattering rates of optical and acoustic phonons in $BC_5$ at T > 100 K can be understood by observing that optical phonons in $BC_5$ have significantly higher frequencies than in materials like silicon. This causes optical phonons in $BC_5$ to scatter by decaying into phonon modes which also have high frequencies. In Fig. 7a we show the dominant contributions to scattering phase space (indicative of the number of scattering channels) of an optical phonon mode with wavevecor $\mathbf{q} = (0.25,0,0)(2\pi/a)$ (where $a$ is the lattice parameter) and frequency $\omega = 1021.6$ cm$^{-1}$. A phonon mode with frequency $\omega$ can scatter either by an absorption process (energy conservation: $\omega + \omega' = \omega''$) or by a decay process (energy conservation: $\omega = \omega' + \omega''$) represented by the first and second delta functions in Equation 2. The $x$-axis of Fig. 7a corresponds to lower of the two frequencies $\omega'$ and $\omega''$ (denoted by $\omega'$ in Fig. 7a) involved in the scattering of above outlined optical phonon mode with frequency $\omega$. The high frequency of the optical phonon (in Fig. 7a) ensures that frequencies of phonons involved in the decay channels are also high (in the range of 400 cm$^{-1}$). It is also clearly visible that decay channels make a large contribution to overall scattering phase space of optical phonons.

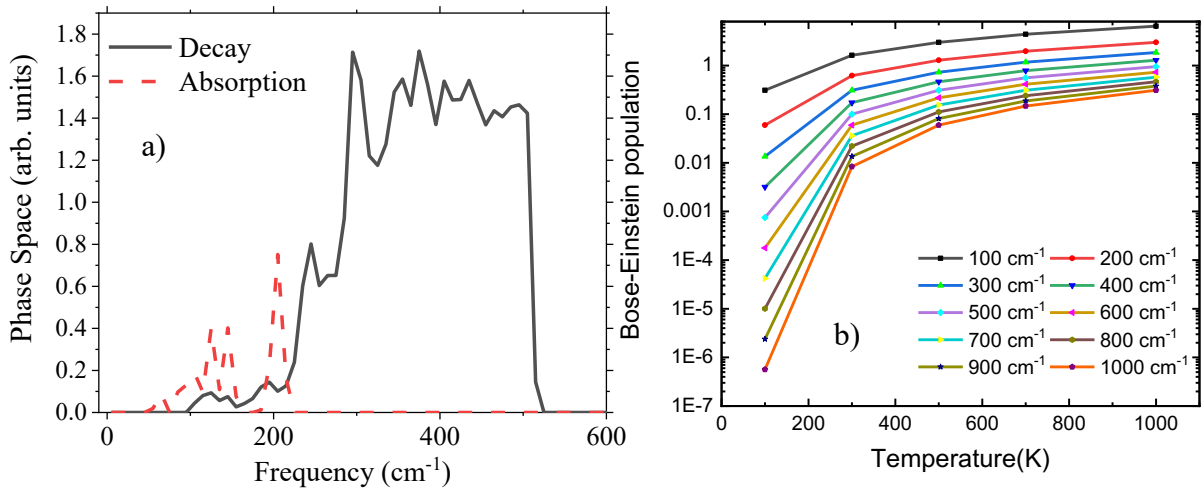

Figure 7: a) Scattering Phase space and b)Bose-Einstein population of phonons in $BC_5$.

The relative insensitivity of optical phonon linewidths to temperature can now be understood by noticing that the scattering rates due to decay processes are proportional to $1 + n_{\lambda'} + n_{\lambda''}$ (second term in Equation 2). The high frequencies of the phonons (~400 cm$^{-1}$ as seen in Fig. 7a) involved in decay of optical phonons causes their populations ($n_{\lambda'}$ and $n_{\lambda''}$) to be remain less than 1.0 even as the temperature increases from 100 K to 800 K (Fig. 7b). The presence of a constant prefactor of 1 in the decay term $1 + n_{\lambda'} + n_{\lambda''}$, then ensures that the overall magnitude of the term, $1 + n_{\lambda'} + n_{\lambda''}$, does not increase significantly over the temperature range of 100 – 800 K, due to the populations, $n_{\lambda'}$ and $n_{\lambda''}$, not exceeding 1.0 over this temperature range. This causes the linewidths of optical phonons to increase slowly with temperature.

For acoustic phonons, however, the dominant scattering mechanism involves absorption scattering channels, which have a population dependence of $n_{\lambda'} - n_{\lambda''}$ (first term in Equation 2). Acoustic phonon scattering rates thus increase in direct proportion to the increase in phonon populations with temperature resulting in a strong increase in linewidths of acoustic phonons. This coupled with only a small increase in linewidths of optical phonons with increase in temperature, causes the linewdiths of the two phonon modes to become comparable at temperatures of 300 K and higher (Fig. 6). These results alongwith high phonon group velocities of optical phonons and a large phonon density of states of optical phonon modes at frequencies $\omega > 500$ cm$^{-1}$ causes optical phonon modes to be a dominant heat carrying channel for the BC$_5$ system.

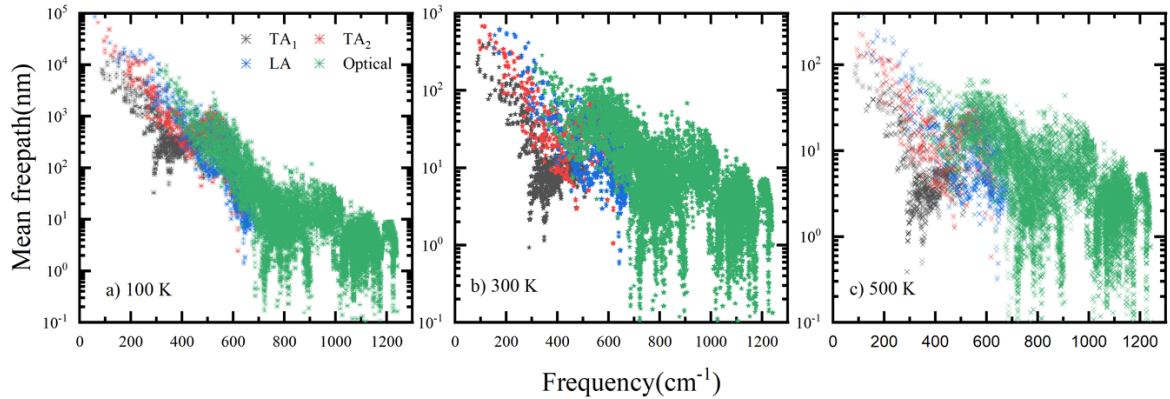

Figure 8: Phonon meanfreepaths in BC$_5$ at a) 100 K b) 300 K and 500 K.

Optical phonon meanfreepaths in BC$_5$ are in the range of nanometers (tens of nanometers) as seen in Fig. 8. The large optical phonon contribution to thermal conductivity coupled with optical phonon meanfreepaths being in the nanometer regime, leads to a large contribution to $k$ in nanoscale regime in BC$_5$. Furthermore, as temperature increases to 300 K and higher, the large increase in acoustic phonon linewidths also causes their meanfreepaths to decrease to nanometers (Fig. 8), further contributing to high nanoscale $k$ in BC$_5$. $k$ contribution from TA, LA and optical phonons for BC$_5$ at nanometer length of L=100 nm in comparison with silicon is discussed in supplementary information(S4).

Though the hardness of BC$_5$ is comparable to diamond, its thermal conductivity is much lower than diamond. The presence of large number of optical phonon branches, results in a dramatic increase in scattering phase space, resulting in phonon scattering rates being significantly larger in BC$_5$ than in diamond (Fig. 9). $k$ contribution from TA, LA and optical phonons for diamond is discussed in supplementary information(S2b).

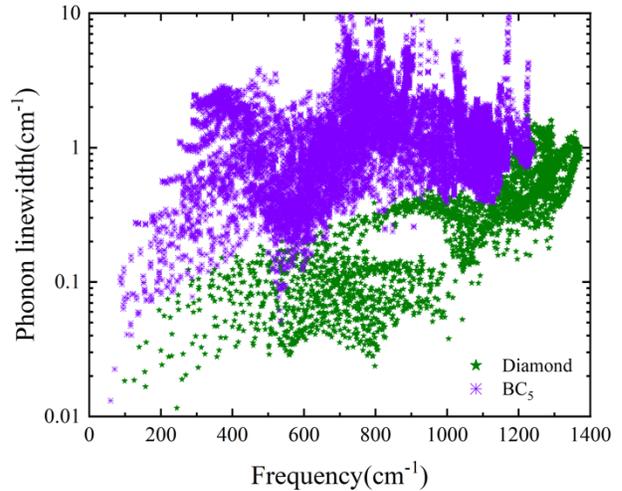

Figure 9: Phonon linewidths of BC$_5$ and Diamond

In summary, using first principles calculations, we analyzed the thermal conductivity of an ultrahard BC$_5$ by solving the Boltzmann transport equation exactly. At room temperature, we report a high thermal conductivity of 169 Wm$^{-1}$K$^{-1}$ for the bulk BC$_5$ and 51 Wm$^{-1}$K$^{-1}$ at the nanometer length scales of L= 50 nm. Ultrahard BC$_5$ will be a promising material for thermal management due to its high thermal conductivity. We also reveal the contributions of optical phonons to overall thermal conductivity to be dominant at high temperatures. At 500 K, optical phonons contribute ~54% to the overall thermal conductivity. This large contribution of optical phonons to overall $k$ was found to be due to comparable group velocities and scattering rates to acoustic phonons at temperatures greater than 300 K. The comparable scattering rates of optical phonons arise from the particular population dependence of decay channels involved in scattering of optical phonons. The large contribution of optical phonons coupled with their meanfreepaths being in the nanometer regime leads to high nanoscale thermal conductivity in $BC_5$. These results may lead to potential applications of $BC_5$ in nanoscale thermal management.

**Conflicts of Interest**

There are no conflicts of interest to declare.

**Acknowledgements**

RM and JG acknowledge support from National Science Foundation CAREER award under Award No. #1847129. We also acknowledge OU Supercomputing Center for Education and Research (OSCER) for providing computing resources for this work.

**Supplementary Information**

**Diamond:**

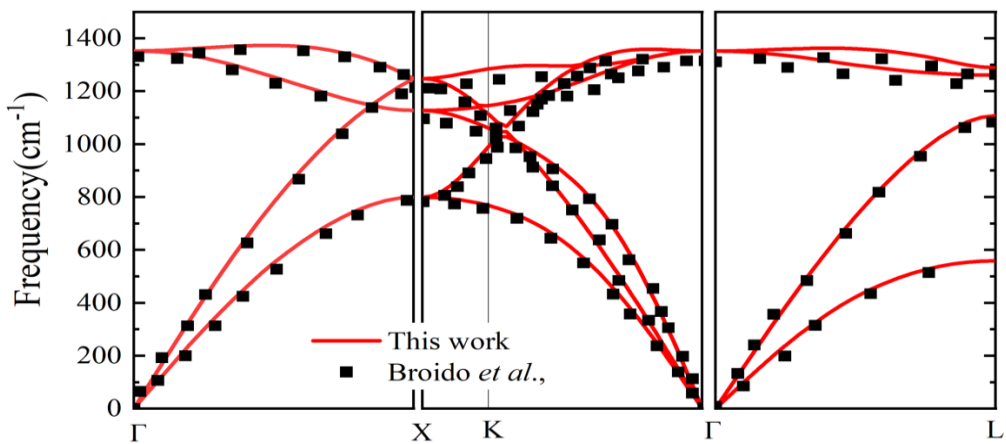

Figure S1: Phonon dispersion for the c_diamond

Figure S1 shows the phonon dispersion for the cubic diamond which are in good agreement with the previous work by Broido *et al.* work. We used the energy cut off of 70 Ry and 8 x 8 x 8 k-point mesh were used to integrate the Brillouin zone. 2nd order force constants were computed on 8 x 8 x 8 q-points and acoustic sum rules were imposed to the 2nd order force constants.

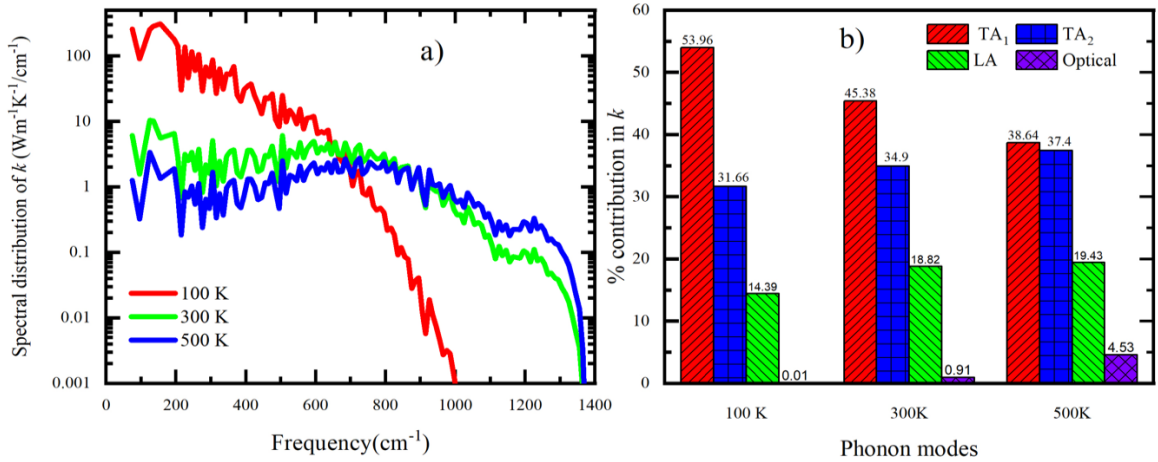

Figure S2a) Spectral distribution and b) thermal conductivity contribution from different phonon branches at 100K, 300 K and 500 K for c_Diamond.

Figure S2b represensts the thermal conductivity contribution by TA, LA and optical phonon modes for the bulk cubic diamond at different temperature using SMA. We can observe that, at 500 K thermal conductivity from the optical phonons are minimal(~4.5 %) compared to 57.3% in $BC_5$.

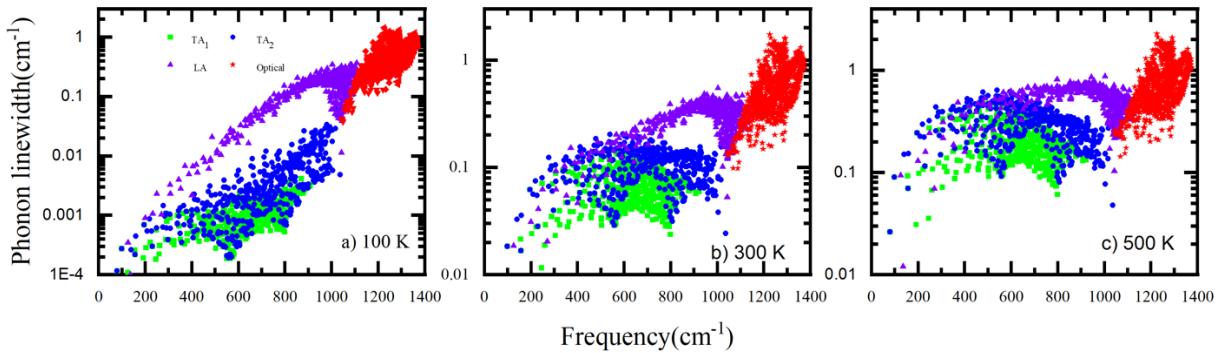

Figure S3: Phonon linewidth for the c_diamond at a)100 K, b)300 K and c)500 K.

Figure S3 shows the phonon linewidth for the cubic diamond at 100 K, 300 K and 500 K. We can observe that, optical phonons has higher scattering rate than the acoustic phonons even at high temperature(500 K).

**Silion**

For silicon , we used the energy cut off of 70 Ry and 8 x 8 x 8 k-point mesh were used to integrate the Brillouin zone. 2$^{nd}$ order force constants were computed on 8 x 8 x 8 q-points and acoustic sum rules were imposed to the 2$^{nd}$ order force constants.

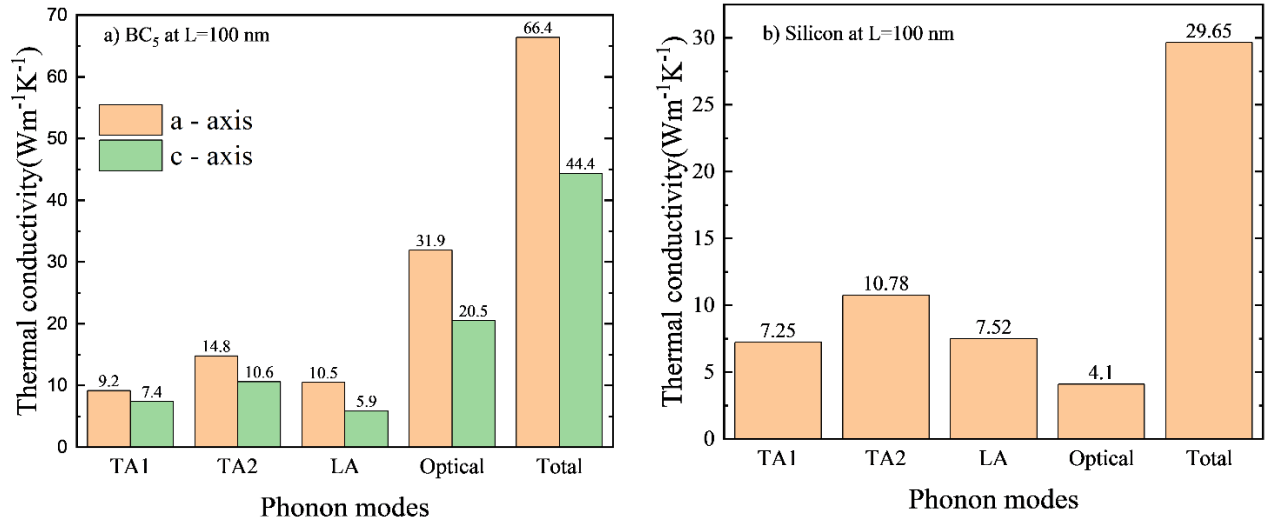

Figure S4 : Thermal conductivity contribution from TA, LA and optical phonons for BC5 and silicon at L=100 nm

Thermal conductivity(SMA) contribution from the TA, LA and optical phonon modes at nanometer length scales of L=100 nm are shown in Fig S4. We also shown the length dependence thermal conductivity of BC5 and silicon in Fig. S5 and we can observe that, at nanometer length scales, thermal conductivity of BC5 is ~120% higher than the silicon(at L=100 nm).

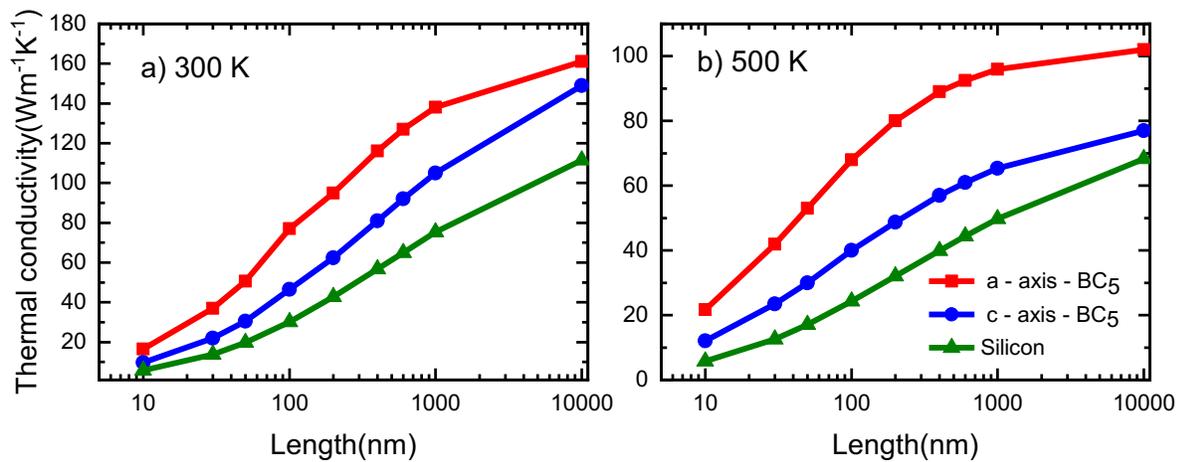

Figure S5: Length dependence thermal conductivity of BC5 and silicon at a) 300 K and 500 K